\documentclass[aps,prb,superscriptaddress,twocolumn,showpacs]{revtex4-1}
\usepackage{amssymb}
\usepackage{epsfig}
\DeclareMathAlphabet{\pazocal}{OMS}{zplm}{m}{n}            
\usepackage{graphicx}
\usepackage{color}
\usepackage{bm}
\usepackage[normalem]{ulem}     
\usepackage{soul}
\usepackage{amsmath}
\usepackage{braket} 
\usepackage{rotating}
\usepackage{multirow} 
\DeclareMathAlphabet{\pazocal}{OMS}{zplm}{m}{n}            
\usepackage{graphicx}
\usepackage{hyperref}

\begin{document}
\title{Electric toroidal dipole order and hidden spin polarization in ferroaxial materials}

\author{Sayantika Bhowal}
\affiliation{Materials Theory, ETH Zurich, Wolfgang-Strasse 27, 8093 Zurich, Switzerland}
\affiliation{Department of Physics, Indian Institute of Technology Bombay, Mumbai 400076, India}

\author{Nicola A. Spaldin}
\affiliation{Materials Theory, ETH Zurich, Wolfgang-Strasse 27, 8093 Zurich, Switzerland}

\date{\today}

\begin{abstract}
We investigate the role of electric toroidal dipoles in the prototypical ferroaxial materials NiTiO$_3$ and K$_2$Zr(PO$_4$)$_2$, which undergo ferroaxial structural phase transitions of order-disorder and displacive type, respectively. Using first-principles electronic structure theory, we compute the evolution across the ferroaxial transitions of the local electric toroidal dipole moments, defined both in terms of the vortices formed by local dipoles, as well as as the cross product of orbital and spin angular momenta. Our calculations confirm that the electric toroidal dipole acts as the order parameter for these ferroaxial transitions and highlight the importance of spin-orbit coupling in generating a non-zero atomic-site electric toroidal dipole moment. We find that, while the ferroaxial phases of NiTiO$_3$ and K$_2$Zr(PO$_4$)$_2$ preserve global inversion symmetry, they contain inversion-symmetry-broken sub-units that generate vortices of local electric dipole moments. In addition to causing the net electric toroidal dipole moment, these vortices induce a hidden spin polarization in the band structure.   
\end{abstract}

\maketitle
\section{Introduction}

\begin{figure}[t]
\includegraphics[width=\columnwidth]{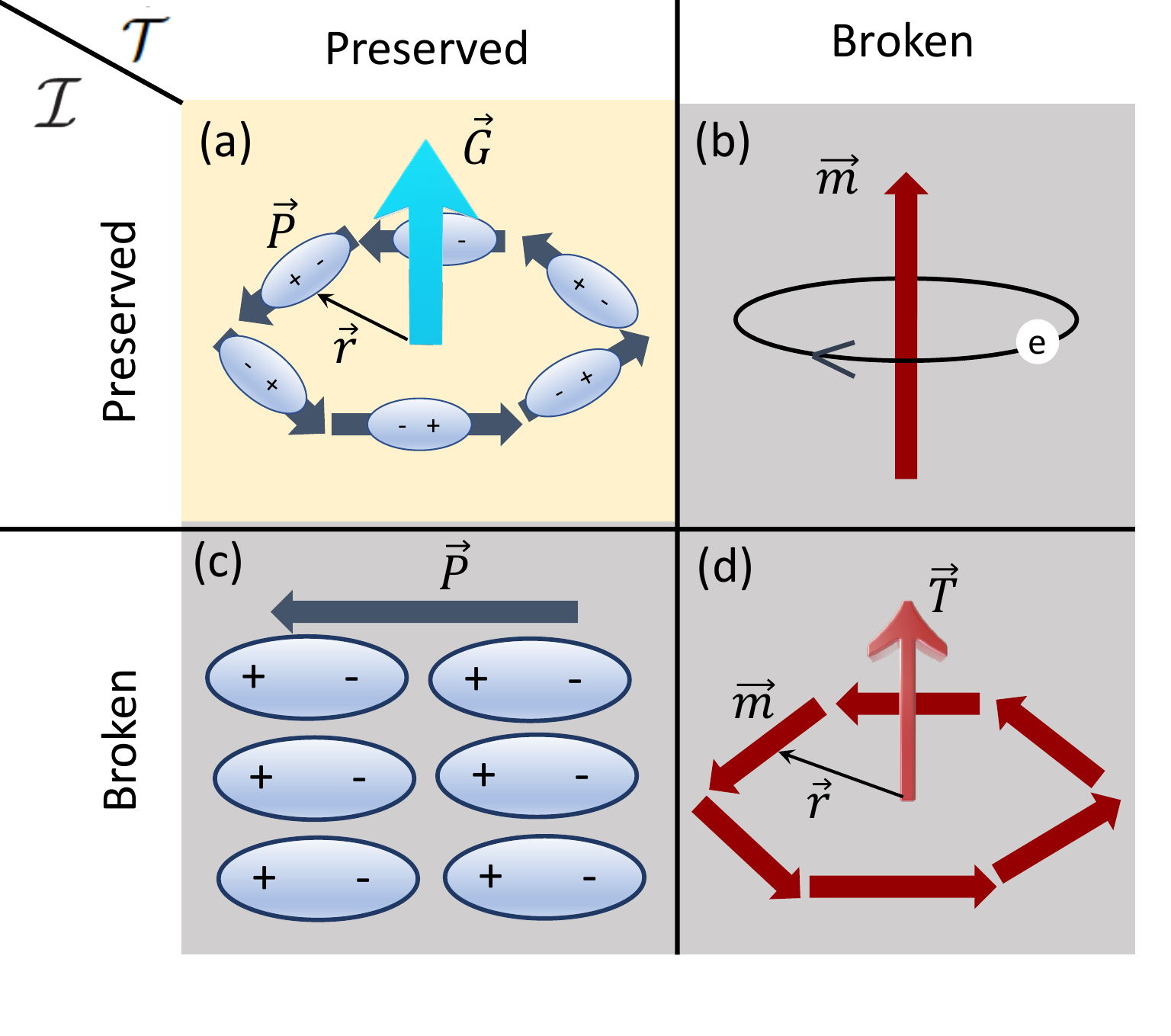}
\caption{An illustration of the four dipolar electromagnetic order parameters with their different symmetries under inversion ($\cal{I}$) and time reversal ($\cal{T}$). These are (a) the electric toroidal dipole moment ($\vec{G}$), which can be generated by a vortex of electric dipole moments and is the primary focus of this work; (b) the magnetic dipole moment ($\vec{m}$); (c) the electric dipole moment ($\vec{P}$); and (d) the magnetoelectric toroidal dipole moment ($\vec{T}$), which can be generated by a vortex of magnetic dipole moments.
}
\label{fig1}
\end{figure}

Symmetry-lowering phase transitions in condensed matter systems give rise to diverse forms of ferroic ordering, which can in turn strongly influence the electromagnetic properties. For example, the breaking of time-reversal ($\cal{T}$) symmetry in ferromagnets leads to ordered magnetic dipoles and spontaneous magnetization, whereas the breaking of inversion ($\cal{I}$) symmetry in ferroelectrics results in a spontaneous electric polarization associated with ordered electric dipoles. The simultaneous breaking of both $\cal{I}$ and $\cal{T}$ symmetries in linear magnetoelectric antiferromagnets, produces a spontaneous magnetoelectric multipolization through the ordering of multipoles such as the magnetoelectric toroidal dipole moment \cite{Spaldin2008, Aken}. Classifying these ordered dipoles according to their symmetries reveals a missing ferroically ordered dipole that breaks neither $\cal{I}$ nor $\cal{T}$ symmetries (see Fig. \ref{fig1}). The electric toroidal dipole (ETD) moment, which has recently gained renewed attention in the context of ferroaxial materials \cite{Hlinka2016, Watanabe2022, Hayashida2021,Hayami2022}, has been proposed as a suitable candidate to fill this gap \cite{Dubovik1990,Prosandeev2006,EdererSpaldin2007}.

The ETD ($\vec G$) is an axial vector that is invariant under both $\cal{T}$ and $\cal{I}$ reversal \cite{Hayami2022}. Classically, it can be constructed from a collection of electric dipole moments ($\vec P$) arranged in a vortex pattern and is expressed as $\vec {G}^{(c)} \sim \sum_i \vec {R}_i \times \vec{P}_i$, where $\vec{R}_i$ signifies the position of the site $i$ hosting the local electric dipole moment $\vec{P}_i$. Recently, an atomic-scale description of ETD has also been obtained \cite{Hoshino2023}, defining it in terms of two different $\cal{T}$-odd vectors, $\vec G \sim \vec l \times \vec s$, where $\vec l$ and $\vec s$ represent the orbital and spin angular momenta, respectively \cite{Kusunose2020, Chikano2021, Hayami2022}.

Unlike the extensively studied magnetic, electric, and magnetoelectric toroidal dipoles, ETDs remained relatively unexplored until recently, since their $\cal{T}$ and $\cal{I}$ symmetries suggest an absence of associated interesting electromagnetic responses. While the natural conjugate field is the inconvenient curl of the electric field $\vec \nabla \times \vec E$ \cite{Dubovik1990,Prosandeev2006}, so-called {\it rotational} responses, described by antisymmetric, off-diagonal tensors in which application of a uniform external field generates the corresponding conjugate physical quantity in the perpendicular direction, have recently been noted \cite{Gopalan2011,Hayami2022, KirikoshiHayami2023, Takeshi2023}. Consequently, optical methods such as electric quadrupole second-harmonic generation and electrogyration (electric field-induced optical rotation) have enabled the spatial resolution of ferroaxial domains \cite{Jin2020, Hayashida2024}. 

While theoretical works \cite{Hayami2022,KirikoshiHayami2023, Takeshi2023} based on symmetry principles and model Hamiltonian calculations have contributed significantly to the fundamental understanding of ETDs, experimental advances \cite{Hayashida2021,Jin2020, Hayashida2024} call for explicit computation of ETDs in real materials. To the best of our knowledge, no computational study on real materials has yet confirmed the symmetry-guided proposal that the ETD serves as the order parameter for the ferroaxial transition. The current manuscript contributes in this regard by quantifying the ETDs in two representative ferroaxial materials, NiTiO$_3$ \cite{Hayashida2021} and K$_2$Zr(PO$_4$)$_2$ \cite{Yamagishi2023}. Both materials undergo a structural phase transition to a ferroaxial phase as the temperature decreases, with NiTiO$_3$ exhibiting an order-disorder transition \cite{Hayashida2024} and K$_2$Zr(PO$_4$)$_2$ undergoing a displacive-type transition \cite{Yamagishi2023}.

Our study provides crucial insights into the nature and role of ETDs in ferroaxial materials. Specifically, we first calculate the quantitative evolution of the atomic-scale ETD, $\vec G \sim \vec l \times \vec s$, across these distinct ferroaxial phase transitions. We find that it is zero in the non-ferroaxial phase and has opposite sign in the two opposite ferroaxial domains, and therefore provides a suitable order parameter for the transition. We note also the crucial role of spin-orbit coupling (SOC) in generating nonzero ETDs. Next, we identify inversion symmetry-breaking structural units (arranged in a manner that retains the global inversion symmetry) within the ferroaxial phases of both materials, containing vortices of local electric dipole moments that yield a non-zero classical $\vec {G}^{(c)} \sim \sum_i \vec {R}_i \times \vec{P}_i$. Again, this definition of the ETD has opposite sign in the two opposite domains and reduces to zero in the non-ferroaxial phases. Finally, we show that these inversion symmetry-breaking units play a crucial role in inducing hidden spin polarization within the bands that we uncover in inversion symmetric NiTiO$_3$ and K$_2$Zr(PO$_4$)$_2$.

The subsequent sections of this manuscript are organized as follows. In Section \ref{structure}, we present a comprehensive analysis of the changes in the crystal structures of NiTiO$_3$ and K$_2$Zr(PO$_4$)$_2$ across the ferroaxial transition. Section \ref{Methods} provides a detailed discussion of the computational methods employed in this study. The results of our calculations are then discussed in Section \ref{results}, where we examine the involvement of electric toroidal dipoles (ETDs) in both the order-disorder and displacement-type ferroaxial phase transitions observed in NiTiO$_3$ and K$_2$Zr(PO$_4$)$_2$, respectively. Additionally, we reveal the presence of hidden spin polarization within the band structures of these materials. Finally, Section \ref{summary} offers a summary of our research work, along with an outlook for further investigations. 

\section{Crystal structure} \label{structure} 

Ferroaxial materials undergo a symmetry-lowering ferroaxial transition characterized by a rotational structural distortion that breaks mirror symmetry while preserving $\cal I$. In the present work, we focus on two ferroaxial materials, NiTiO$_3$ and K$_2$Zr(PO$_4$)$_2$.

\begin{figure}[t]
\includegraphics[width=\columnwidth]{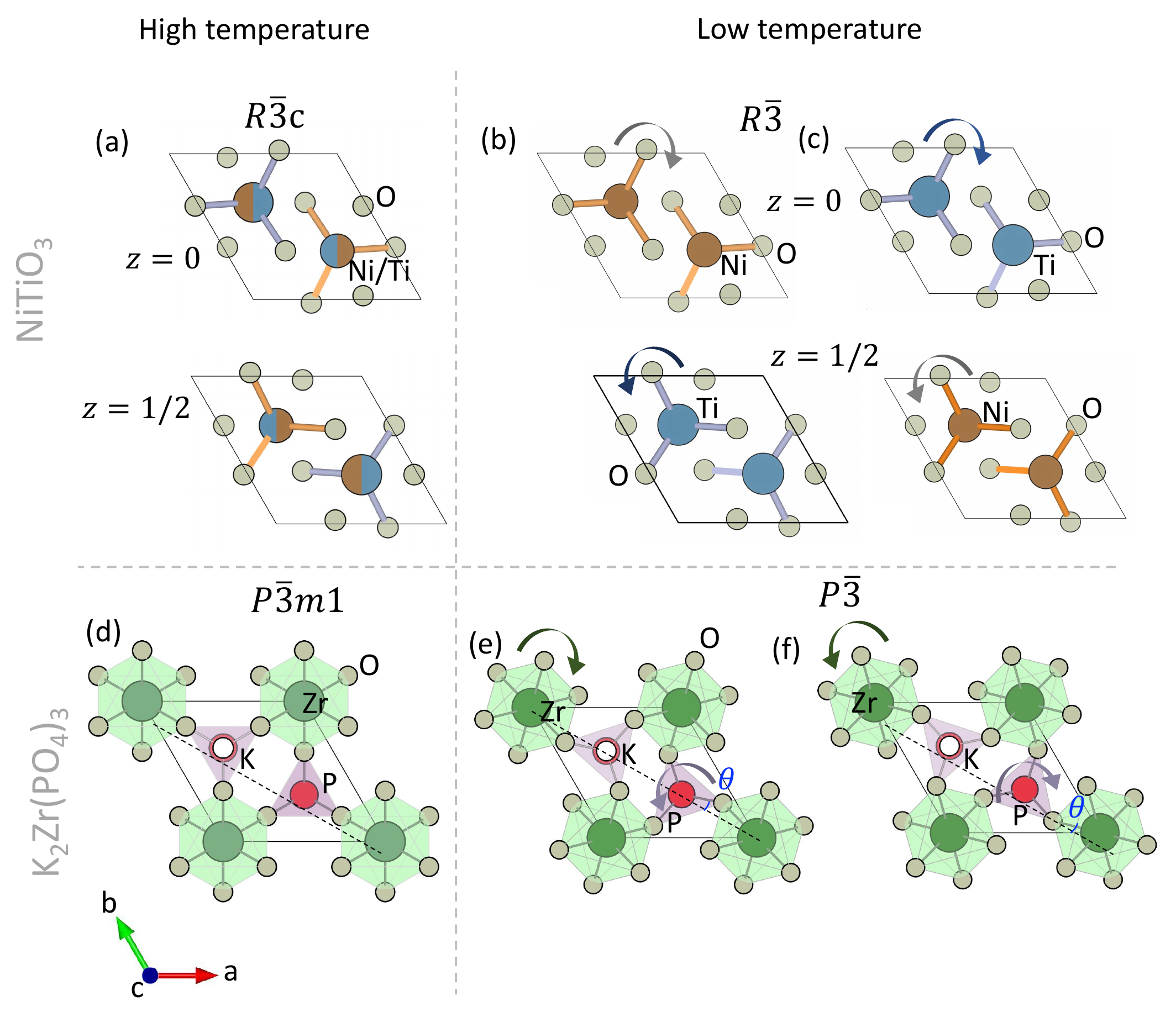}
\caption{Crystal structures of NiTiO$_3$ and K$_2$Zr(PO$_4$)$_3$. (a) The disordered non-ferroaxial $R\bar{3}c$ crystal structure of NiTiO$_3$ at high-temperature. (b-c) The ferroaxial domains of NiTiO$_3$ below the ferroaxial transition temperature. The opposite rotation of the transition metal-oxygen networks at two different planes for each domains is apparent from the figure. The planes of the Ni and Ti atoms are switched between the two domains. (d) The high-temperature non-ferroaxial $P\bar{3}m1$ structure and (e-f) the low-temperature ferroaxial ($P\bar{3}$) domains of K$_2$Zr(PO$_4$)$_3$. The rotation by an angle of $\theta$ of the PO$_4$ tetrahedra with respect to the non-ferroaxial structure is indicated in the figure.   
}
\label{fig2}
\end{figure}

NiTiO$_3$ undergoes a structural phase transition at approximately $T_c \approx 1560$ K, transitioning from a disordered corundum structure (nonferroaxial $R\bar{3}c$) with a random distribution of Ni$^{2+}$ and Ti$^{4+}$ ions to an ordered ilmenite structure (ferroaxial $R\bar{3}$) as the temperature decreases \cite{Lerch1992}. The crystal structures of NiTiO$_3$ at high and low temperatures are shown in Fig. \ref{fig2} (a-c). The conventional unit cell of NiTiO$_3$, both below and above $T_c$, contains 6 formula units. In the ordered $R\bar{3}$ structure below $T_c$, the Ni$^{2+}$ and Ti$^{4+}$ ions are arranged in alternating planes (Ni-Ti-Ni-Ti) $\perp c$, with each ion surrounded by six oxygen atoms, forming NiO$_6$ and TiO$_6$ octahedral networks that share their faces along the $c$ direction (See Fig. \ref{fig4}a). Notably, the rotational distortions of the surrounding oxygen atoms are opposite for consecutive layers (see Fig. \ref{fig2}) in both low- and high-temperature structures. In the high temperature $R\bar{3}c$ structure, the cations are randomly distributed (see Fig. \ref{fig2}a), and the measured cation-oxygen bond lengths are identical for consecutive layers, so the average rotations in adjacent layers are equal and opposite. In contrast, in the low-temperature structure, the Ni-O and Ti-O bond lengths differ, so the opposite rotations are of  different size. Consequently, two structural domains form depending on the arrangement of cations in different planes (e.g., Ni-Ti-Ni-Ti and Ti-Ni-Ti-Ni) as depicted in Fig. \ref{fig2} (b-c). 

In contrast to the order-disorder transition observed in NiTiO$_3$, the glaserite-type compound K$_2$Zr(PO$_4$)$_2$ undergoes a displacement-type ferroaxial transition around 700 K, from a high-temperature $P\bar{3}m1$ structure to a low-temperature $P\bar{3}$ structure \cite{Yamagishi2023} as depicted in Fig. \ref{fig2} (d-f). The crystal structure of K$_2$Zr(PO$_4$)$_2$ consists of ZrO$_6$ octahedral and PO$_4$ tetrahedral units that share corners. Below the transition temperature, these corner-shared networks of ZrO$_6$ and PO$_4$ units undergo a rotation with a rotational angle $\theta$ of 17.6$^\circ$, resulting in ferroaxial order (see Fig. \ref{fig2} (d-f)). Consequently, depending on the direction of rotation, two opposite ferroaxial domains are formed, as depicted in Fig. \ref{fig2} (e-f).

\section{computational methods} \label{Methods}

The electronic structures of NiTiO$_3$ and K$_2$Zr(PO$_4$)$_2$ were computed using the plane-wave based projector augmented wave (PAW) method \cite{Bloch1994, Kresse1999} within the local density approximation (LDA) as implemented in the Vienna ab initio simulation package (VASP) \cite{Kresse1993, Kresse1996}. To achieve self-consistency, a kinetic energy cut off of 550 eV was used for the wavefunctions and a $k$ mesh of $8\times8\times3$ is used to sample the Brillouin zone (BZ) of NiTiO$_3$. For K$_2$Zr(PO$_4$)$_2$, the chosen kinetic energy cut off and the $k$ mesh were respectively 550 eV and $10\times10\times6$. Calculations were performed both in the presence and absence of SOC. The PAW potentials employed in the calculations were O ([He]$2s^22p^4$), P ([Ne]$3s^23p^3$), K\_sv ([Ne]$3s^23p^64s^1$), Zr\_sv ([Ar]$4s^24p^64d^35s^1$), Ni ([Ar]$3d^94s^1$), and Ti\_sv ([Ne]$3s^23p^63d^34s^1$). Atomic relaxations were performed until the Hellman-Feynman forces on each atom were less than 0.01 eV/\AA. The spin textures in momentum space were computed using the open-source Python library PyProcar \cite{Herath2020,Lang2024}.

For the computation of the local electric toroidal dipole and electric dipole moments, multipole calculations were carried out using the methodology outlined in Refs. (\cite{Cricchio2009, Bultmark2009, Cricchio2010}). In this approach, the density matrix $\rho_{lm,l'm'}$ is decomposed into irreducible (IR) spherical tensor components $w^{k p r}_t$. Here, $k, p, r$ denote the spatial index, spin index (with $p = 0$ for charge and $p = 1$ for magnetic multipoles), and the rank of the tensor ($r \in { | k - p |, | k - p | + 1, \hdots , k + p }$), while $t \in { -r, -r + 1, \hdots , r }$ labels the component of the tensor. The spherical tensor $w^{1 1 1}_t \propto (\vec l \times \vec s)$ \cite{Laan1995} and, hence, it is a measure of the atomic-site ETD component \cite{Hayami2022,Hoshino2023}. The atomic-site ETD being inversion symmetric, the tensor moments have contributions from $l+l'=$ even terms in the density matrix $\rho_{lm,l'm'}$. For the atomic-site odd-parity electric dipole moment, the tensor moment $w^{1 0 1}_t$ has only contributions from $l\ne l'$ terms. The tensor $w^{1 0 1}_t$ is a measure of the atomic-site electric dipole component. Since it has odd parity, it only has contributions from $l\ne l'$ terms.

\section{Results and Discussions} \label{results}

In this section, we present and analyze our computational results. Our main findings are that: (i) both $\sum_i \vec{R}_i \times \vec{P}_i$ and $\vec{l} \times \vec{s}$ serve as measures of ETD, and (ii) the ETD is a suitable order parameter for the ferroaxial transition, since it is zero in the high-temperature non-ferroaxial phase and increases in magnitude as the rotational distortion increases. We illustrate the important role of ETD in mediating order-disorder and displacement-type ferroaxial transitions by considering NiTiO$_3$ and K$_2$Zr(PO$_4$)$_2$ as examples. In addition we reveal a hidden spin polarization in these inversion symmetric materials, which is a consequence of the ETD and in turn a signature of the ferroaxial order.

\subsection{ETD and the order-disorder ferroaxial phase transition in NiTiO$_3$} \label{NTO}

We start with the electronic structure of NiTiO$_3$. While NiTiO$_3$ is known to undergo an antiferromagnetic phase transition below $T_N = 22.5$ K \cite{Stickler1967, Harada2016, Dey2020}, in the present work, since our focus is on the ETDs, we do not consider the effect of magnetism in our calculations. Hence, our computational results correspond to the regime $T_c > T > T_N$ in which the system does not have any long-range magnetic ordering. Within this regime our LDA+SOC calculations yield metallic behavior, with the Fermi energy primarily dominated by the Ni-$d$ states, hybridized with the Ti-$d$ and O-$p$ states.    

\begin{figure}[t]
\includegraphics[width=\columnwidth]{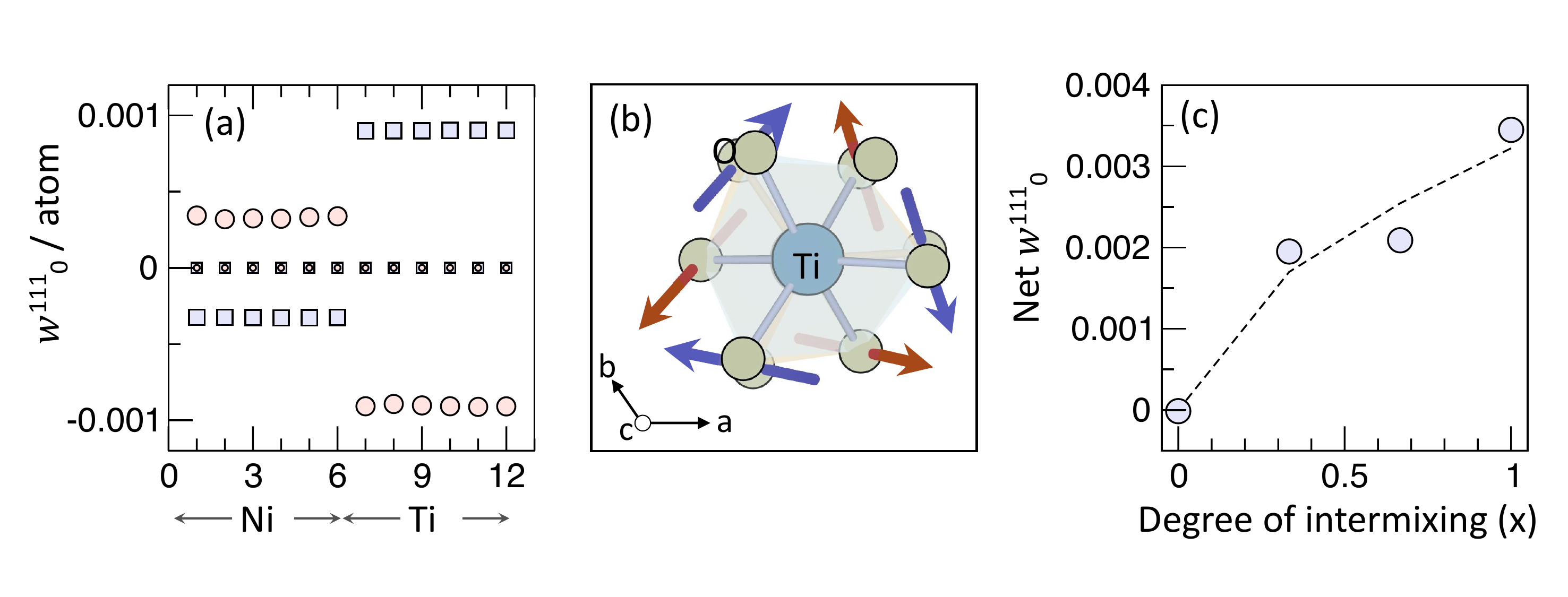}
\caption{ETD in NiTiO$_3$. (a) Computed $w^{111}_0$ multipole which is a measure of electric toroidal dipole moment component $G_z$ for each transition metal atom (Ni/Ti) in the unit cell for the two ferroaxial domains of NiTiO$_3$, both in the presence (large squares and circles) and absence (small squares and circles) of SOC. The circles and squares represent the computed results for the ferroaxial $G_z^{+}$ (see Fig. \ref{fig2}b) and $G_z^{-}$ (see Fig. \ref{fig2}c) domains, respectively. (b) Top view of the computed $w^{101}$ multipole (which indicates the local electric dipole moments) at the O atoms, surrounding the Ni (blue arrows) and Ti (red arrows) atoms, indicating opposite vortices of electric dipole moments around the Ni and Ti atoms. (c) The variation of the net $w^{111}_0$ multipole (i.e., the net $G_z$) as a function of the degree of intermixing $x$ (see text for details). The dashed line provides a guide to the eye.}
\label{fig3}
\end{figure}

We compute the atomic-site ETDs for the two structural domains of NiTiO$_3$, depicted in Fig. \ref{fig2} (b-c). All calculations are carried out with the relaxed crystal structure of NiTiO$_3$ (see section \ref{Methods} for details). Our calculations show that the $w^{111}_0$ spherical tensor component, which is $\propto \vec l \times \vec s$, has a non-zero value at the Ni and Ti atoms, indicating the presence of local $G_z$ ETD components in NiTiO$_3$. The results of our calculations are shown in Fig. \ref{fig3} (a). As seen from this plot, the ETDs at the 6 Ni (Ti) sites have the same sign, consistent with the totally symmetric $A_{1g}$ IR representation of the $C_{3i}$ point group symmetry of NiTiO$_3$. However, between the Ni and Ti atoms, the ETDs have opposite signs. Notably, the magnitude of the ETDs at the Ni atoms differs from that at the Ti atoms, resulting in a net ETD in the unit cell of NiTiO$_3$. In other words, the ferroaxial phase of NiTiO$_3$ has a ``ferri-type" ordering of the ETDs between the Ni and Ti atoms. Furthermore, we find that the ETDs at the Ni and Ti atoms reverse their signs for the opposite structural domain (Fig. \ref{fig3}a). Consequently, the net ETD in the unit cell also reverses its sign. Thus the two ferroaxial domains are characterized by a net ETD of opposite signs, suggesting it to be a good order parameter for the ferroaxial phase.

The existence of $G_z$ and its opposite sign at the Ni and Ti atoms are consistent with the rotational distortion and its reversal between the layers containing Ni and those containing Ti in ordered NiTiO$_3$, as discussed in section \ref{structure} and depicted in Fig. \ref{fig2} (b-c). Also, the different Ni-O and Ti-O bond lengths explain the different magnitudes of ETD at the Ni and Ti atomic sites. The sign reversal of ETD for the opposite ferroaxial domain also follows from the opposite octahedral rotations around the Ni and Ti atoms between the two structural domains as evident from Fig. \ref{fig2} (b-c). 

\begin{figure}[t]
\includegraphics[width=\columnwidth]{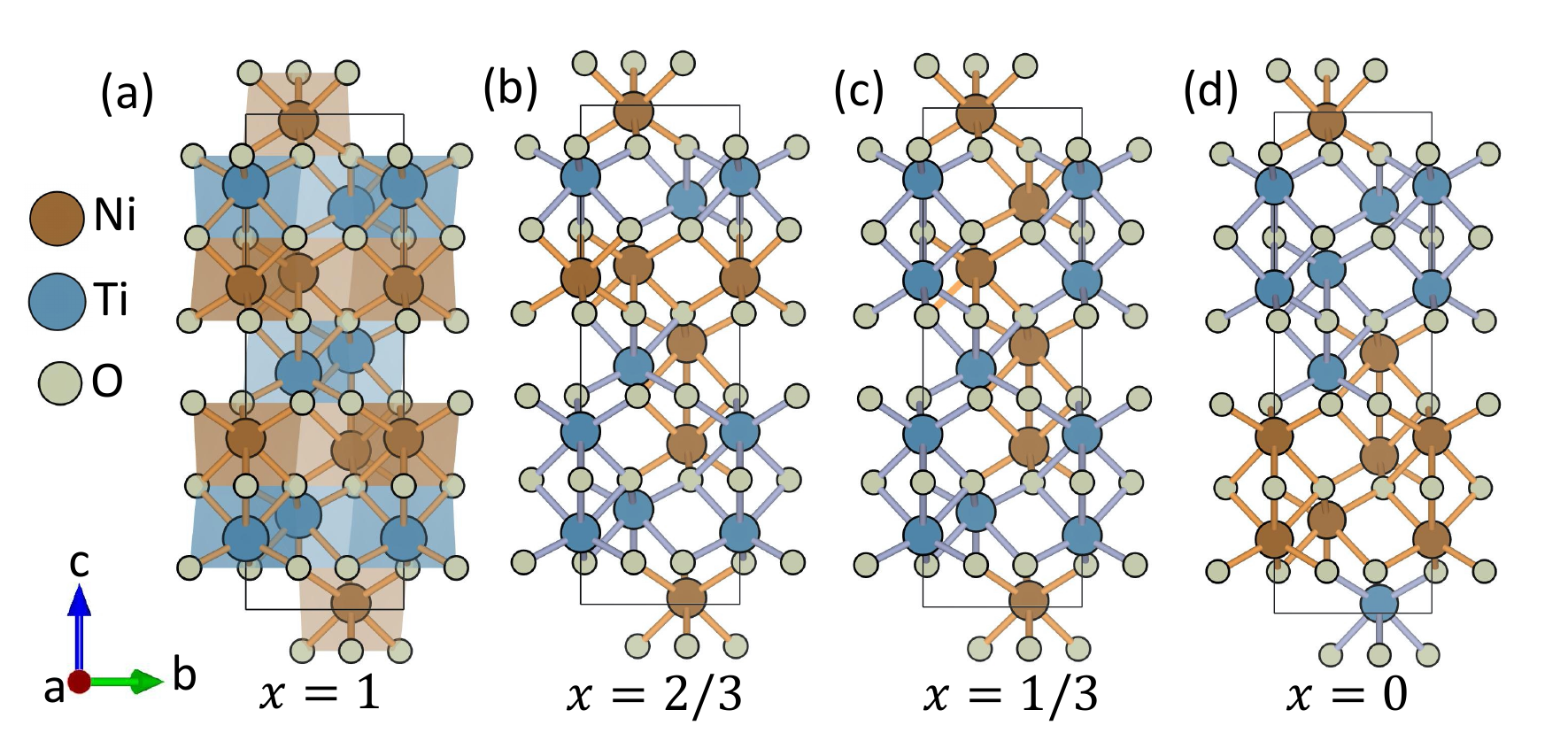}
\caption{The considered structures with different degrees of intermixing (a) The ideal low-temperature structure with $x=1$, (b) $x=2/3$, (c) $x=1/3$, and (d) $x=0$, used for the computation of the net $G_z$ plotted in Fig. \ref{fig3}(c).}
\label{fig4}
\end{figure}

{\it Effect of disorder}- Based on this understanding, we intuitively expect that if the atoms are disordered, the ETDs will average to zero. To further verify our understanding and to study the effect of disorder on the ETDs in NiTiO$_3$, we construct a few model structures in which selected Ni and Ti atoms are interchanged. To describe these structures, we define the degree of intermixing $x$ as,
\begin{equation}
x = \frac{|P_{\rm Ni}-P_{\rm Ti}|}{P_{\rm Ni}+P_{\rm Ti}}, 
\end{equation}
where $P_{\rm Ni}$ ($P_{\rm Ti}$) is the probability that all Ni sites are occupied by Ni (Ti). In the ordered structure, all 6 Ni sites are occupied by Ni, so that $P_{\rm Ni}=1, P_{\rm Ti} = 0$ and hence, $x = 1$ (see Fig. \ref{fig4}a). To mimic the effect of cation inter-mixing in the disordered structure, we  construct structures by exchanging one of the Ni (Ti) atoms with Ti (Ni), which leads to $P_{\rm Ni}=5/6$ and $P_{\rm Ti} = 1/6$, resulting in $x = 2/3$. Similarly, structures corresponding to $x = 1/3,$ and 0 are also constructed by respectively interchanging 2 and 3 Ni and Ti atoms. The considered structures with different $x$ values are shown in Fig. \ref{fig4}. 

It is important to highlight that our $x = 0$ structure is constructed to emulate the high temperature disordered $R\bar3c$ structure of NiTiO$_3$, in which all the transition metal-oxygen bond lengths are equal in magnitude. We therefore used the reported crystal structure (both lattice parameters and internal coordinates of the atoms) at high-temperature in Ref. \cite{Boysen1995} for our calculations of the $x = 0$ structure. For the other disordered structures ($x = 1/3$ and $2/3$) we further relaxed all the internal coordinates keeping the lattice parameters fixed to experimental values, before computing the ETDs. We note that our choice of structure for a given intermixing parameter $x$ is not unique, as the Ni atom can, in principle, occupy any of the six available Ti sites, and our use of periodic boundary conditions imposes a periodicity that does not exist in a real disordered structure. Nonetheless, our considered representative structures effectively capture the essential physics of cation intermixing in disordered structures on the ETDs.

The sums of the computed local $w^{111}_0 \propto \vec l \times \vec s$ multipoles for each of the structures shown in Fig. \ref{fig4} are depicted in Fig. \ref{fig3} (c). As we can see from the plot, the net $w^{111}_0$ multipole in the unit cell decreases in magnitude with increasing degree of intermixing, and finally, vanishes to zero for $x=0$, indicating that the ETD is a suitable order parameter for the order-disorder transition in NiTiO$_3$. 
The physical understanding of this behavior follows from two important effects of the disorder. First of all, when a Ni site is occupied by a Ti atom, the sign of the ETD at that Ti atom changes relative to its sign on the Ti site in the ordered structure. This is because the sense of rotation is different for the Ni and Ti sites in the ordered structure. In addition, moving one of the Ti atoms to a Ni site also affects the corresponding transition metal-oxygen bond lengths.
 We find that disorder tends to reduce the differences between the Ni-O and Ti-O bond lengths, decreasing, thereby, the difference in magnitude of $w^{111}_0$ at the Ni and Ti atomic sites. Therefore, although the atomic-site $w^{111}_0$ persists at the Ni and Ti atoms in the disordered structures, the magnitude of the net $w^{111}_0$ in the unit cell decreases with decreasing $x$. For the $x=0$ structure, we use the experimentally reported identical TM-O bond lengths, explaining the vanishing net $w^{111}_0$ in the unit cell.

{\it Effect of spin-orbit interaction} Next, we investigate the effect of SOC in determining the magnitude of the $w^{111}_0$ multipole, by artificially switching off the SOC in our calculations. We find that the $w^{111}_0$ multipole at the Ni and Ti atoms vanishes as the SOC is turned off (see Fig. \ref{fig3}a), highlighting the crucial role of SOC in determining the atomic-site ETD. Physically, we understand this behavior from the fact that in a nonmagnetic system, local spin degrees of freedom are only active in the presence of SOC, and hence $\vec G = \vec l \times \vec s$ has a non-zero value only in the presence of spin-orbit interaction.

Finally, to assess the validity of the alternative classical definition of ETDs at the Ni and Ti atomic sites, we compute the $w^{101}$ spherical tensor, which is a measure of the local electric dipole moment, at the O atoms surrounding the Ni and Ti atoms. The computed local electric dipole moment vectors, dictated by the $w^{101}_t$ multipole ($t=-1, 0, 1$ correspond to the $y, z,$ and $x$ components of the electric dipole moment vector respectively), are shown in top view (i.e. in the plane $\perp c$) in Fig. \ref{fig3} (b). The vector representations of the $w^{101}_t$ multipoles at the O atoms form a vortex-like arrangement around the transition metal ions, with the vortices being clockwise and counter-clockwise around the Ni and Ti atoms respectively. Since the $w^{101}_t$ multipole is a measure of electric dipole moments, this confirms the presence of vortices of electric dipole moment around the Ni and Ti atoms in the plane $\perp c$, which, in turn, indicates the existence of $G_z$ according to the classical definition $\vec {G}^{(c)} \sim \sum_i \vec {R}_i \times \vec{P}_i$, as stated before.

\subsection{ETD and the displacement-type ferroaxial phase transition of K$_2$Zr(PO$_4$)$_2$}

Having established that the ETD, described by both $ \vec l \times \vec s$ and $\sum_i \vec {R}_i \times \vec{P}_i$, is the order parameter for the ferroaxial transition in  NiTiO$_3$, we now turn to the displacive-type ferroaxial transition in K$_2$Zr(PO$_4$)$_2$. 
The transformation from the non-ferroaxial $P\bar3m1$ to ferroaxial $P\bar3$ space group is driven by a pure rotational distortion ${\cal R}_z$ around the $z$ axis, corresponding to the $A_{2g}$ IR of the $-3m1$ point group symmetry of the high-temperature structure. We generate intermediate crystal structures between non-ferroaxial $P\bar3m1$ and ferroaxial $P\bar3$ of both domains (see Fig. \ref{fig2} (d-e)) by varying the size of the distortion mode, identified using the ISODISTORT program \cite{Stokes, Campbell2006}. 

\begin{figure}[t]
\includegraphics[width=\columnwidth]{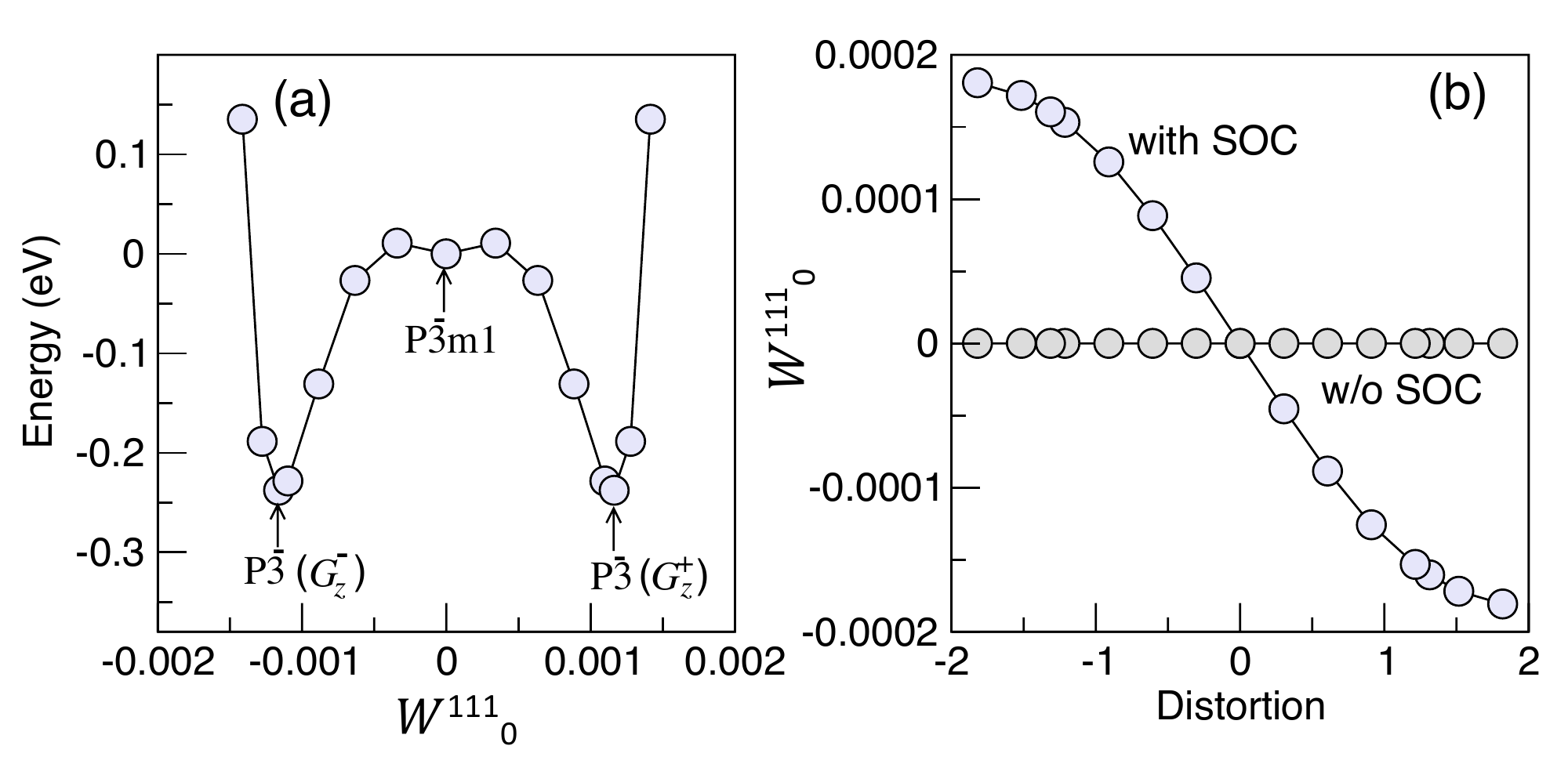}
\caption{ETD as an order parameter for the ferroaxial phase transition of K$_2$Zr(PO$_4$)$_2$. (a) The variation of the computed energy of K$_2$Zr(PO$_4$)$_2$ as a function of the electric toroidal dipole (ETD) moment. The energy of the high-temperature $P\bar{3}m1$ structure (see Fig. \ref{fig2}d) is set to zero. The $G_z^{+}$ and $G_z^{-}$ ferroaxial domains correspond to the structures shown in Fig. \ref{fig2}e  and f respectively. (b) The variation of the total ETD, computed by summing the $p$-$s$ contribution to the $w^{111}_0$ multipole at every atomic site in the unit cell,  as a function of the rotational distortion ${\cal R}_z$ both in the presence and absence of SOC.}
\end{figure}

\begin{figure}[t]
\includegraphics[width=\columnwidth]{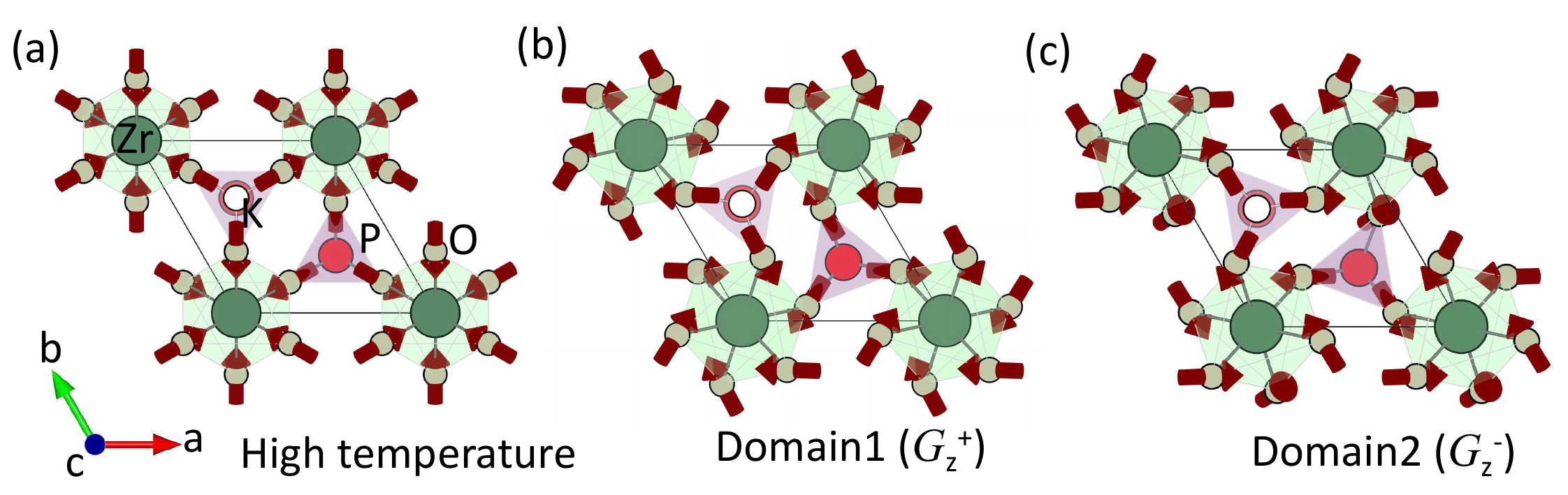}
\caption{The computed $w^{101}$ multipole (which indicates the local electric dipole moments) at the O atoms surrounding the Zr and P atoms for: (a) the high-temperature structure, (b) the low-temperature ferroaxial domain $G_z^{+}$, and (c) the low-temperature ferroaxial domain $G_z^{-}$.}
\label{fig5b}
\end{figure}

We then perform self-consistent calculations to compute the total energies and multipole calculations to evaluate the atomic-site $w^{111}$ multipoles for these intermediate crystal structures in addition to the end-point structures. We find that, similarly to the case of NiTiO$_3$, the opposite ferroaxial domains of K$_2$Zr(PO$_4$)$_2$ are also characterized by a net non-zero $z$ component of the $w^{111}$ multipole of the same magnitude but opposite signs, indicating the ETD as the order parameter. In contrast to the case of NiTiO$_3$, however, $w^{111}_0$ multipoles vanish at every atomic site in the high-temperature non-ferroaxial $P\bar3m1$ structure of  K$_2$Zr(PO$_4$)$_2$. To further illustrate that the ETD is the order parameter for the ferroaxial phase, we show in Fig. \ref{fig5a} (a) the variation of the total energy as a function of the computed net $w^{111}_0$ multipoles in K$_2$Zr(PO$_4$)$_2$. The total energy exhibits the well-known $\phi^4$ profile characteristic of second-order phase transitions within Landau theory, confirming the ETD as the order parameter for the ferroaxial phase transition in K$_2$Zr(PO$_4$)$_2$. We note that the total energy corresponding to the zero multipole case is a bit lower which maybe attributed to the different crystal symmetry, identified in the DFT computation.  The variation of the total $w^{111}_0$ multipole (i.e., the sum of the $p$-$s$ contribution to the $w^{111}_0$ multipole at every atomic site in the unit cell) as a function of the rotational distortion, as shown in Fig. \ref{fig5a} (b), depicts the increase in the ETD with an increase in the distortion as well as its sign reversal upon reversal of the sense of rotation. 

To verify the alternative definition of ETD, we further compute the $w^{101}_t$ multipoles (representing the electric dipole) at the oxygen atoms surrounding the Zr atom for both high-temperature and low-temperature structures and depict the computed multipoles as vectors in Fig. \ref{fig5b} (a-c). The arrangement of the local $w^{101}_t$ multipoles around the Zr atom in the high-temperature non-ferroaxial structure shows that the multipoles are all directed towards the central Zr atom. Thus, the ETD moment, representing a vortex arrangement of the electric dipole moment dictated by the $w^{101}_t$ multipoles, vanishes. For the low-temperature ferroaxial structures, however, the $w^{101}_t$ multipole moments have a vortex-like structure with clockwise and counter-clockwise rotations for the two opposite domains, verifying that it is a good definition of ETD and is the order parameter for the ferroaxial transition.

\subsection{Hidden spin polarization}

Both NiTiO$_3$ and K$_2$Zr(PO$_4$)$_2$ preserve inversion and time-reversal symmetries, leading to doubly degenerate bands of opposite spin polarization over the entire BZ. However, the existence of vortices of local electric dipole moments forming the ETDs in ferroaxial materials, demonstrated in the previous sections, suggests that the inversion symmetry is broken locally in the crystal structure. This, in turn, leads to a hidden spin polarization \cite{Zhang2014} in these materials, which we discuss in this section.

\begin{figure}[t]
\includegraphics[width=\columnwidth]{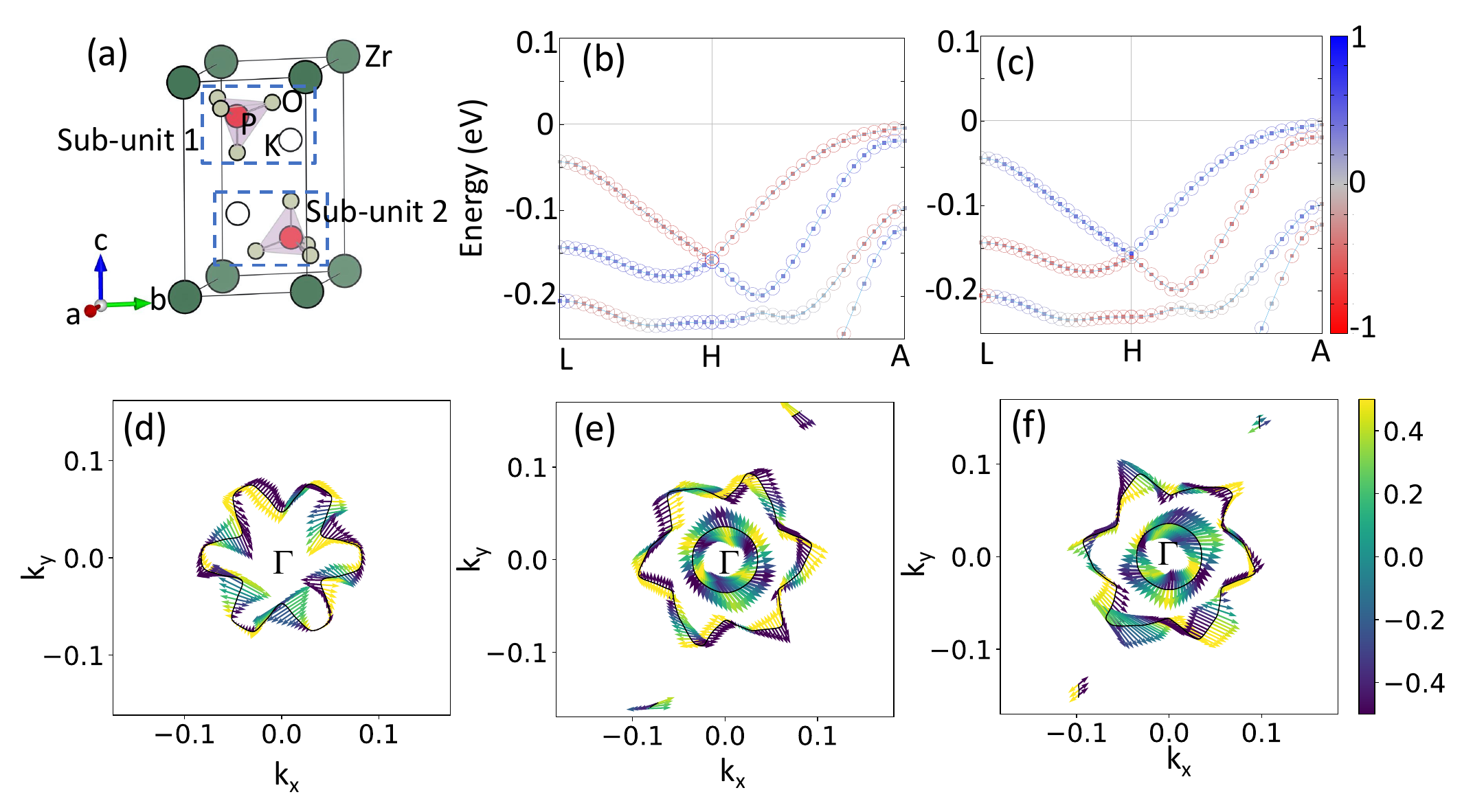}
\caption{Hidden spin polarization in K$_2$Zr(PO$_4$)$_2$. (a) Crystal structure of K$_2$Zr(PO$_4$)$_2$, highlighting the broken inversion symmetric units, labeled as sub-unit 1 and sub-unit 2. Spin $S_z$ projection (indicated by the color bar) from the atoms in (b) sub-unit 1 and (c) sub-unit 2 for each of the individual bands of a doubly degenerate pair in the band structure of K$_2$Zr(PO$_4$)$_2$. The spin texture in the $k_x$-$k_y$ plane of the BZ of K$_2$Zr(PO$_4$)$_2$ at an energy 0.164 eV below the Fermi energy in its (d) high-temperature non-ferroaxial phase, and ferroaxial domains (e) $G_z^+$ and (f) $G_z^-$.}
\label{fig6}
\end{figure}

To confirm the local inversion symmetry breaking of the crystal structure, we first analyze the Wyckoff site symmetries of the atoms in ferroaxial K$_2$Zr(PO$_4$)$_2$. While the Zr atom occupies an inversion-symmetric site (note that this is consistent with the atomic-site ETD on Zr, since the ETD preserves inversion symmetry), all other atoms (i.e., K, P, and O atoms) occupy Wyckoff sites that do not have inversion symmetry. More importantly, a careful inspection of the crystal structure, as depicted in Fig. \ref{fig6}a, shows that the unit cell consists of two sub-units of K(PO$_4$). While each of these sub-units individually breaks inversion symmetry, the presence of both units restores the global inversion symmetry of the structure. 

To understand the consequence of these inversion symmetry-breaking units, we compute the band structure of K$_2$Zr(PO$_4$)$_2$ in the presence of SOC. As expected, bands of opposite spin-polarization are degenerate in energy due to the presence of global inversion and time-reversal symmetries. We find that each of these bands has equal contributions from both sub-units. More interestingly, the spin projection of the individual sub-units, as depicted in Fig. \ref{fig6} (b-c), shows that each sub-unit gives rise to spin-split bands which are degenerate with another set of spin-split bands of opposite spin polarization originating from the other sub-unit. Thus, the inversion symmetry breaking of the sub-units gives rise to a hidden spin polarization of the bands in the globally doubly degenerate bands.

We note that a hidden spin polarization is also present in the high-temperature structure of K$_2$Zr(PO$_4$)$_2$, due to the presence of local inversion symmetry breaking units. During the ferroaxial transition, however, locally the mirror symmetry is also broken. For example, the point group symmetry of Wyckoff site $2d$ corresponding to atomic positions of P, K, and one of the O atoms changes from $C_{3v} \rightarrow C_3$. The breaking of mirror symmetry affects the hidden spin texture in the material. This is evident in the computed spin texture in the $k_x$-$k_y$ plane of the BZ of K$_2$Zr(PO$_4$)$_2$ corresponding to the high-temperature structure compared to the two ferroaxial domains as shown in Fig. \ref{fig6} (d-f) respectively. Note that since K$_2$Zr(PO$_4$)$_2$ is an insulator, spin textures are computed for the valence bands and, more specifically, the spin textures in Fig. \ref{fig6} (d-f) correspond to the bands at an energy 0.164 eV below the Fermi energy. As shown in Fig. \ref{fig6} (e-f), the spin texture exhibits a Rashba-like \cite{Rashba1960,BychkovRashba,PekarRashba1964} form, favoring the alignment of spins perpendicular to the momentum direction, resulting in a $k$-space magnetoelectric (ME) toroidal moment (a head-to-tail spin arrangement) \cite{Bhowal2022}. However, it is not a pure $k$-space ME toroidal moment, as evidenced by the slight canting of the spins. This canting is likely due to the Dresselhaus effect \cite{Dresselhaus1955}, which is allowed by the local Wyckoff site symmetries. Therefore, the spin texture is a combination of local Rashba and Dresselhaus effects for the two ferroaxial domains. Interestingly, as seen from Fig. \ref{fig6} (e-f), the spin textures form vortices around the $\Gamma$ point with the sense of rotation being opposite for the opposite ferroaxial domains analogous to the corresponding oppositely rotated vortices of the local electric dipole moments. For the high-temperature structure, although the spin texture exists, it does not have a vortex-like arrangement due to the presence of $\sigma_v$ mirror symmetry, which also leads to vanishing ETD in the high-temperature structure.

\begin{figure}[t]
\includegraphics[width=\columnwidth]{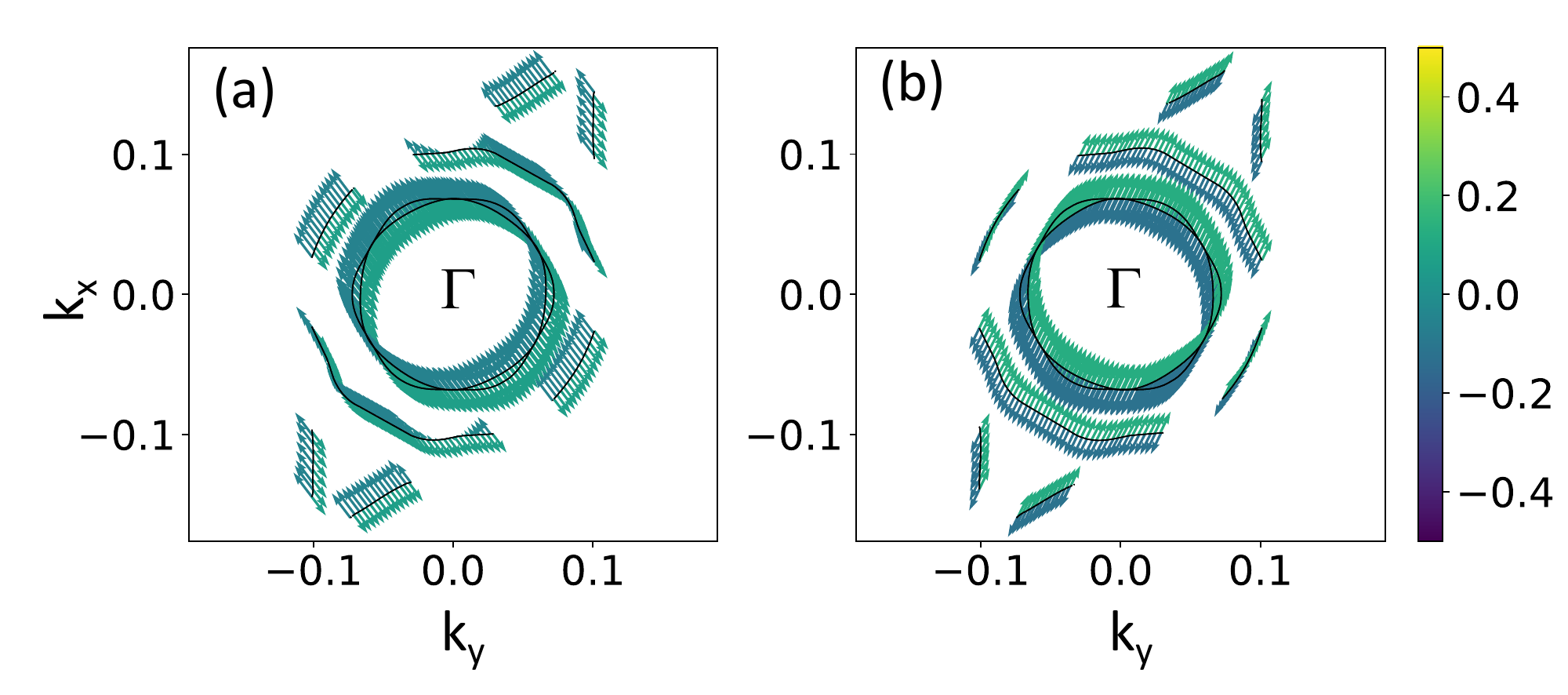}
\caption{The hidden spin texture in the $k_x$-$k_y$ plane of the BZ of NiTiO$_3$ at the Fermi energy in the ferroaxial domains (a) $G_z^+$ and (b) $G_z^-$. Color map depicts the spin projection.}
\label{fig7}
\end{figure}

A similar hidden spin texture also exists in the ferroaxial phase of NiTiO$_3$.  The computed spin textures for the two ferroaxial domains of NiTiO$_3$ are shown in Fig. \ref{fig7} (a) and (b), depicting again the rotation of the spin texture between the two ferroaxial domains. 

We note that similar hidden spin polarization has been predicted previously in other inversion symmetric systems, such as $\alpha$-SnTe, NaCaBi, and LaOBiS$_2$ \cite{Zhang2014}. Our work confirms its existence in ferroaxial materials and points out the intriguing correlation between the ETD, the order parameter for the ferroaxial transition, and the corresponding spin texture in reciprocal space.  
Interestingly, as discussed in Ref. \cite{Zhang2014}, the compensated hidden spin polarization of electronic bands in the BZ of ferroaxial materials can be manipulated by symmetry-breaking perturbations such as an external electric field. By making the sub-unit layers inequivalent, the external perturbations will lead to an uncompensated spin polarization, revealing thereby the hidden spin texture of the unperturbed system.

\section{Summary and outlook} \label{summary}

To summarize, we provide, for the first time, a quantitative description of the ETD in real materials, from the perspective of both the traditional classical description, $\vec {G}^{(c)} \sim \sum_i \vec {R}_i \times \vec{P}_i$ as well as the recently proposed quantum-mechanical atomic-site expression, $\vec G \sim \vec l \times \vec s$. 

Taking the examples of NiTiO$_3$ and K$_2$Zr(PO$_4$)$_2$, we show how the ETD evolves across the ferroaxial transition and find that the ETD is both an order parameter for the transition and an indicator of the type -- order-disorder or displacive -- of transition. For the displacive ferroaxial transition of K$_2$Zr(PO$_4$)$_2$, the atomic site ETDs vanish in the non-ferroaxial structure, whereas for the order-disorder type transition of NiTiO$_3$, the atomic-site ETDs remain non-zero in the non-ferroaxial structure, with cation disorder leading to the cancellation of these atomic-site ETDs and a net zero value. 

While neither the ETD nor the ferroaxial order break inversion symmetry, we find that intra-unit cell inversion symmetry breaking, which nevertheless cancels at the unit cell level, is an essential ingredient in generating the vortices of electric dipole moment. This symmetry breaking combines with the SOC that is crucial for the existence of the atomic-site ETDs to generate a hidden spin texture in the BZ of ferroaxial materials. This could be revealed by extracting single or odd numbers of layers of locally broken inversion symmetry from bulk ferroaxial materials, or by using external perturbations such as an electric field.
Since inversion symmetry breaking gives rise to orbital moments in reciprocal space even without SOC, the BZ of ferroaxial materials may also host hidden orbital textures in addition to the spin textures discussed in the present work. The interplay between the relative orientation of the orbital and spin moments in reciprocal space, and the  atomic-site ETDs $\sim (\vec l \times \vec s)$ is an interesting  topic for future work. Another open question is the role of the ETD in the recently observed polar skyrmions\cite{Das2019}

Finally, we note that ferroaxial materials maybe thought of as ``protochiral",since breaking the last remaining mirror plane in ferroaxial materials leads to the emergence of chiral crystal structures \cite{HayashidaKimura2021,Zheng2024}. Indeed, the inversion symmetry-breaking monopole of ETDs has been proposed as the order parameter for chirality \cite{Oiwa2022}, and the BZs of chiral materials are known to exhibit spin-polarized bands \cite{Winkler2024} due to the imbalance created by mirror symmetry breaking between the otherwise compensated sub-units in ferroaxial materials. In this context, it will be intriguing to explore the effects of an external electric field perpendicular to the mirror plane of a ferroaxial material or examine the surface parallel to the mirror plane, as both break the symmetries required for chirality. We hope our work will inspire further theoretical studies and experimental investigations in this area in the near future.

\section*{Acknowledgements}
 The authors thank Carl Romao, Roland Winkler and Uli Zuelicke for stimulating discussions.
NAS and SB were supported by the ERC under the EU’s Horizon 2020 Research and Innovation Programme grant No 810451 and by the ETH Zurich. SB thanks IIT Bombay for financial support. Computational resources were provided by ETH Zurich's Euler cluster, and the Swiss National Supercomputing Centre, project ID eth3.

\bibliography{LNPO}
\end{document}